\documentclass[a4paper]{article}

\usepackage{graphicx}
\usepackage{onecolpceurws}

\usepackage{amsfonts}
\usepackage{dsfont}

\usepackage{mathtools}
\usepackage{amssymb}
\usepackage{amsmath}
\usepackage{amsthm}
\usepackage{epstopdf}
\usepackage{enumitem} 
\usepackage{caption}
\usepackage{float}
\usepackage{subcaption}
\usepackage{epsfig}
\usepackage{multirow}
\usepackage[all]{xy}
\usepackage{mdframed}
\usepackage{multirow}
\usepackage{url}
\usepackage{lipsum}
\newtheorem{definition}{Definition}[section]

\newtheorem{theorem}{Theorem}[section]

\title{Formal Modeling of Robotic Cell Injection Systems in Higher-order Logic}

\author{
Adnan Rashid and Osman Hasan \\ School of Electrical Engineering and Computer Science (SEECS) \\
                National University of Sciences and Technology (NUST) \\
                Islamabad, Pakistan \\ \{adnan.rashid,osman.hasan\}@seecs.nust.edu.pk
}

\institution{}

\begin{document}
\maketitle

\begin{abstract}
Robotic cell injection is used for automatically delivering substances into a cell and is an integral component of drug development, genetic engineering and many other areas of cell biology. Traditionally, the correctness of functionality of these systems is ascertained using paper-and-pencil proof and computer simulation methods. However, the paper based proofs can be human-error prone and the simulation provides an incomplete analysis due to its sampling based nature and the inability to capture continuous behaviors in computer based models. Model checking has been recently advocated for the analysis of cell injection systems as well. However, it involves the discretization of the differential equations that are used for modeling the dynamics of the system and thus compromises on the completeness of the analysis as well. In this paper, we propose to use higher-order-logic theorem proving for the modeling and analysis of the dynamical behaviour of the robotic cell injection systems. The high expressiveness of the underlying logic allows us to capture the continuous details of the model in their true form. Then, the model can be analyzed using deductive reasoning within the sound core of a proof assistant.
\end{abstract}
\vskip 32pt

\section{Introduction}\label{SEC:Intro}
Robotic cell injection systems are used to automatically and precisely insert small amounts of substances, such as molecules and genes, into cells during various gene injection~\cite{kuncova2004challenges}, drug development~\cite{nakayama1998new}, intracytoplasmic sperm injection (ISCI)~\cite{yanagida1999usefulness} and in-vitro fertilization (IVF)~\cite{sun2002biological}.  The most critical factor in these systems is the precision and accuracy of the injection force~\cite{huang2009visual} as a slight excessive force may damage the membrane of the cell~\cite{huang2006visual} or an insufficient force may not be able to pierce the cell~\cite{faroque2016virtual}.
Moreover, these robotic systems consist of many sub-components, like injection manipulator, digital cameras, sensors and microscope optics~\cite{huang2009visual}, and a controlled movement of these fundamental components is also quite vital for the functionality of the overall system.

In order to attain the above-mentioned objectives, the robotic cell injection systems need to be carefully designed and analyzed. For this purpose, the behavior of a robotic cell injection system's movements has to be modeled using the coordinate frames corresponding to the orientations of its various components, i.e., the injection manipulator, cameras and images. Similarly, we need to capture the motion planning of the injection pipette in terms of force control algorithms, such as the contact-space-impedance force control~\cite{sun1997modeling,huang2009visual} and the image-based torque controller~\cite{huang2006visual}. These models are then analyzed to ensure the desired behavior using paper-and-pencil and simulation techniques. However, the manual analytical analysis is prone to human error and also is not scalable for analyzing complex robotic cell injection systems. Similarly, due to the continuous nature of the analysis and the limited amount of computational resources, the system is analyzed for a certain number of test cases only in simulation and thus the absolute accuracy cannot be achieved. Thus, the above-mentioned traditional techniques cannot be relied upon as they are either error prone or incomplete, which may lead to an undetected error in the analysis that may in turn lead to disastrous consequences given the safety-critical nature of robotic cell injection systems.

Formal methods~\cite{hasan2015formal} are computer-based mathematical analysis techniques that can overcome the above-mentioned inaccuracies. Primarily, these techniques involve the development of a mathematical model of a system and verification of its properties using computer-based mathematical reasoning.
Sardar et al.~\cite{sardar2017towards} recently used probabilistic modeling checking~\cite{clarke1999model}, i.e., a state-based formal method, to formally analyze the robotic cell injection systems. However, their methodology involves the discretization of the differential equations that model the dynamics of these systems, which compromises the accuracy of the corresponding analysis. Moreover, the analysis also suffers from the inherent state-space explosion problem~\cite{clarke2012model}. Higher-order-logic theorem proving~\cite{harrison2009handbook} is an interactive verification technique  that can overcome these limitations. It primarily involves the mathematical modeling of the system based on higher-order logic and verification of its properties based on deductive reasoning. Given the high expressiveness of higher-order logic, it can truly capture the behavior of the differential equations, which is not possible in model checking based analysis.

In this paper, we propose to use the higher-order-logic theorem proving to formally model and analyze the robotic cell injection systems~\cite{huang2006visual} using the HOL Light proof assistant~\cite{harrison1996hol}.
The main motivation for the selection of HOL Light is the availability of reasoning support for real calculus~\cite{hol_light2017realanalysis}, multivariate calculus~\cite{hol_light2017multivariate}, vectors~\cite{hol_light2017multivariate} and matrices~\cite{hol_light2017multivariate}, which are some of the foremost requirements for formally analyzing robotic cell injection systems. We use these foundations to formally model the camera, stage and image coordinates and formal verification of their interrelationships in HOL Light. Similarly, we also formally modeled the dynamics of two degrees of freedom (DOF) motion stage using a system of differential equations and the formal verification of their solutions.

The rest of the paper is organized as follows: Section~\ref{SUBSEC:Mult_cal_theories} provides an introduction about the multivariate calculus theories of HOL Light that we build upon to model the robotic cell injection system. We provide an overview about the robotic cell injection system in Section~\ref{SUBSEC:rob_cell_inj_sys}. Section~\ref{SEC:form_analy_cell_inj_sys} presents the formalization of robotic cell injection system.
Finally, Section~\ref{SEC:conclusion} concludes the paper.

\section{Multivariable Calculus Theories in HOL Light} \label{SUBSEC:Mult_cal_theories}

A $\mathds{N}$-dimensional vector in HOL Light is modeled as a $\mathds{R^N}$ column matrix with each of its element representing a real number~\cite{harrison2013hol}. All of the vector arithmetics are thus done using matrix manipulations. Similarly, all theorems of HOL Light multivariable calculus theories are verified for functions with an arbitrary data-type $\mathds{R^N} \rightarrow \mathds{R^M}$.

We explain some of the frequently used HOL Light functions in the proposed formalization as follows:

\begin{mdframed}
\begin{definition}
\label{DEF:cx_and_ii}
\emph{Vector} \\{
\textup{\texttt{$\vdash$ $\forall$ l.\ vector l = (lambda i.\ EL (i - 1) l)
}}}
\end{definition}
\end{mdframed}

\noindent The function $\mathtt{vector}$ takes an arbitrary list \texttt{l} : $\alpha\ \mathtt{list}$ and results into a vector having each component of data-type $\mathds{\alpha}$. It uses the HOL Light function $\mathtt{EL \ n \ L}$, which extracts the $n^{th}$ element of a list $\texttt{L}$. Here, the \texttt{lambda} operator in HOL is used for constructing a vector based on its components~\cite{harrison2013hol}.

\begin{mdframed}
\begin{definition}
\label{DEF:exp_ccos_csine}
\emph{Real Cosine and  Real Sine} \\{
\textup{\texttt{$\vdash$ $\forall$ x.\ cos x = Re (ccos (Cx x)) \\
$\mathtt{}$$\vdash$ $\forall$ x.\ sin x = Re (csin (Cx x))
}}}
\end{definition}
\end{mdframed}

The functions $\texttt{cos}:\mathds{R} \rightarrow \mathds{R}$
and $\mathtt{sin}:\mathds{R} \rightarrow \mathds{R}$ in HOL Light represent the real cosine and real sine~\cite{hol_light2017transcendentals}, respectively.
These functions are modeled in HOL Light based on the complex cosine \texttt{ccos} : $\mathds{R}^2 \rightarrow \mathds{R}^2$ and complex sine \texttt{csin} : $\mathds{R}^2 \rightarrow \mathds{R}^2$ functions, respectively.

\begin{mdframed}
\begin{definition}
\label{DEF:vector_derivative}
\emph{Real Derivative} \\
{
\textup{\texttt{$\vdash$ $\forall$ f x.\ real\_derivative f x = (@f$\mathtt{'}$.\ (f has\_real\_derivative f$\mathtt{'}$) (atreal x))
}}}
\end{definition}
\end{mdframed}

The function $\mathtt{real\_derivative}$ represents the derivative of a real-valued function and is defined using the Hilbert choice operator $\texttt{@}$ in the functional form. It accepts a real-valued function $\texttt{f} : \mathds{R} \rightarrow \mathds{R}$ and a real number $\texttt{x}$, which is the point at which $\texttt{f}$ has to be differentiated, and returns a variable of data-type $\mathds{R}$, which is the differential of $\texttt{f}$ at $\texttt{x}$. The function $\mathtt{has\_real\_derivative}$ defines the same relationship in the relational form.

We build upon the above-mentioned fundamental functions of multivariable calculus to formally analyze the robotic cell injection system in Section~\ref{SEC:form_analy_cell_inj_sys} of the paper.


\section{Robotic Cell Injection Systems} \label{SUBSEC:rob_cell_inj_sys}

\begin{figure}[!ht]
\centering
\scalebox{0.95}
{\hspace*{-0.4cm} \includegraphics[trim={5.0 0.4cm 5.0 0.4cm},clip]{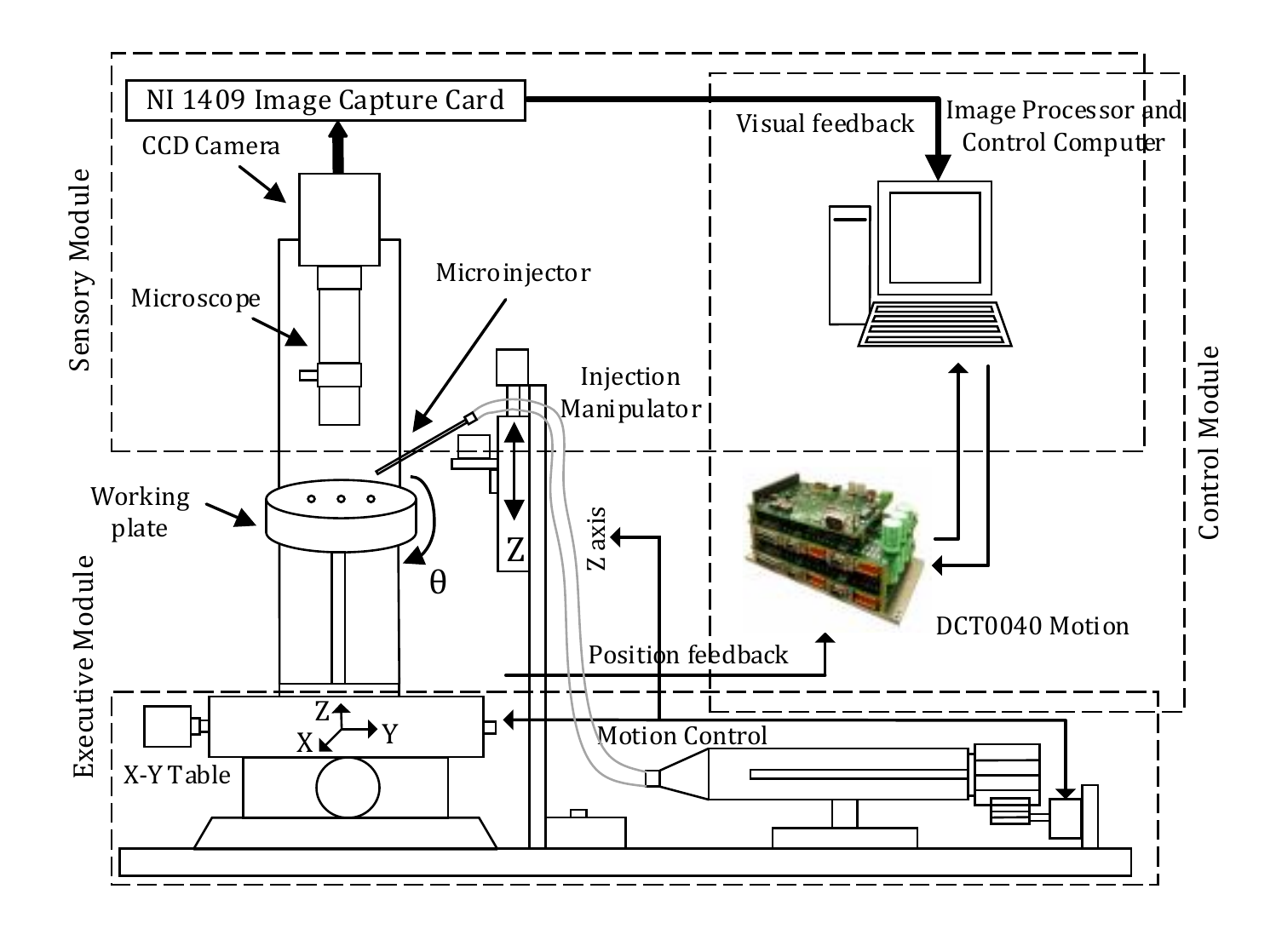}}
\caption{Robotic Cell Injection Systems}
\label{FIG:rob_cell_injec}
\end{figure}

A typical robotic cell injection system is composed of three modules, namely executive, sensory and control modules as depicted in Figure~\ref{FIG:rob_cell_injec}. The executive module comprises of working plate, positioning table and the injection manipulator. The working plate is mounted on the positioning table ($XY\theta$-axis) and thus holds the cells that need to be injected. Similarly, the injection manipulator is mounted on $Z$-axis as shown in Figure~\ref{FIG:rob_cell_injec}.

\begin{figure}[!ht]
\centering
\scalebox{0.75}
{\includegraphics[trim={5.0 0.4cm 5.0 0.4cm},clip]{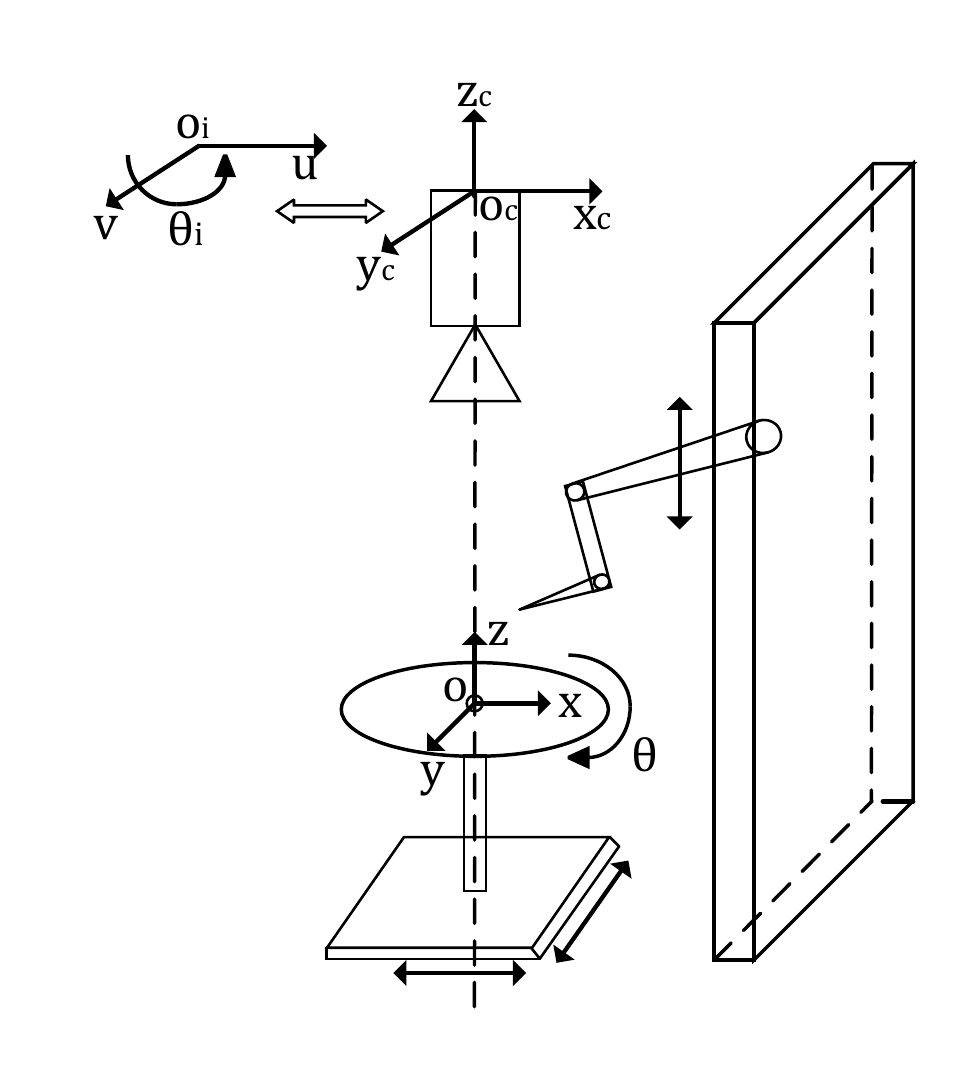}}
\caption{Configuration of the Robotic Cell Injection Systems}
\label{FIG:config_rob_cell_injec}
\end{figure}

The sensory module consists of a vision system that has four components, namely charged coupled device (CCD) camera, processing card, peripheral component interconnect (PCI) image capture and an optical microscope. A PCI image capture alongside a CCD camera is used to capture the process of cell injection. The control module includes a DCT0040 motion control system and a host computer. The configuration of a robotic cell injection system is depicted in Figure~\ref{FIG:config_rob_cell_injec}. The stage (table and working plate) coordinate frame is represented as the axis $o-xyz$, where the origin of these coordinates, i.e., $o$ represents the center of the working plate and the optical axis of the microscope is along the component $z$ of the axis. Similarly, the camera coordinate frame is represented by the axis $o_c-x_cy_cz_c$, where the origin $o_c$ represents the center of the microscope. The axis $o_i-uv$ represents the coordinate frame in image plane with $o_i$ representing its origin and the axis $uv$ is perpendicular to the optical axis.

\section{Formal Modeling of Robotic Cell Injection System}\label{SEC:form_analy_cell_inj_sys}

This section presents the higher-order-logic based formal modeling of the robotic cell injection system. To facilitate the understanding of the paper for a non-HOL user, we use the standard mathematical notations for describing the proposed formalization rather than the HOL Light notations. However, the readers, interested in viewing the exact HOL Light formalization, can find the source code of our formalization at~\cite{adnan17robcellinjsystp}. We consider $2$-DOF to capture the dynamics of the robotic cell injection system. The stage, camera and image coordinates are two-dimensional coordinates, which are formalized in HOL Light as:


\begin{mdframed}
\begin{flushleft}
\begin{definition}
\label{DEF:two_dim_coor}
\emph{Two-dimensional Coordinates} \\{  
\vspace*{0.2cm}
\textup{\texttt{$\vdash$ $\forall$ x y t.\ two\_dim\_coordinates x y t = $\begin{bmatrix}\texttt{x(t)} \\ \texttt{y(t)}\end{bmatrix}$
}}}
\end{definition}
\end{flushleft}
\end{mdframed}

\noindent where \texttt{x} : $\mathds{R} \rightarrow \mathds{R}$ and \texttt{y} : $\mathds{R} \rightarrow \mathds{R}$ modeling the respective axes and \texttt{t} is a variable representing the time.

Next, we formalize the two-dimensional displacement vector between the origins of the stage coordinate frame ($o-xyz$) and the camera coordinate frame ($o_c-x_xy_cz_c$), and the rotation matrix from the stage coordinate frame to the camera coordinate frame as:


\begin{mdframed}
\begin{flushleft}
\begin{definition}
\label{DEF:rot_mat_n_disp_vec}
\emph{Displacement Vector and Rotation Matrix} \\{   
\vspace*{0.1cm}
\textup{\texttt{$\vdash$ $\forall$ dx dy.\ displace\_vector dx dy =  $\begin{bmatrix}\texttt{dx} \\ \texttt{dy}\end{bmatrix}$  \\ \vspace*{0.1cm}
$\mathtt{}$$\vdash$ $\forall$ alpha.\ rotat\_matrix alpha = 
$\begin{bmatrix}\texttt{cos (alpha)} & \hspace*{0.3cm} \texttt{sin (alpha)} \\ \texttt{-sin (alpha)} & \hspace*{0.3cm} \texttt{cos (alpha)} \end{bmatrix}$
}}}
\end{definition}
\end{flushleft}
\end{mdframed}


The verification of the relationship between the camera, image and stage coordinate frames provides the reliable working of the robotic cell injection system as it ascertains the accuracy of the orientation and movement of its various modules, i.e., camera, microscope, stage frame and the injection manipulator. Firstly, we verify the relationship between camera and stage coordinates:


\begin{mdframed}
\begin{flushleft}
\begin{theorem}
\label{THM:rel_camera_stage}
\emph{Relationship Between Camera and Stage Coordinate Frames} \\{
\textup{\texttt{$\vdash$ $\forall$ xc yc x y alpha dx dy t.\  \\ 
$\mathtt{}$ \hspace*{0.0cm} \textbf{[A1]:}\ 0 < dx $\wedge$ \textbf{[A2]:}\ 0 < dy    \\ \vspace*{0.1cm}
$\mathtt{}$ \hspace*{0.8cm} $\Rightarrow$ \Big(relat\_camera\_stage\_coordinates xc yc x y alpha dx dy t $\Leftrightarrow$   \\ \vspace*{0.1cm}
$\mathtt{}$ \hspace*{2.5cm}  $\begin{bmatrix}\texttt{xc(t)} \\ \texttt{yc(t)}\end{bmatrix}$ = $\begin{bmatrix}\texttt{x(t) $\ast$ cos (alpha) + y(t) $\ast$ sin (alpha) + dx} \\ \texttt{- x(t) $\ast$ sin (alpha) + y(t) $\ast$ cos (alpha) + dy}\end{bmatrix}$\Bigg)
}}}
\end{theorem}
\end{flushleft}
\end{mdframed}

\vspace*{0.2cm}

\noindent where the function \texttt{relat\_camera\_stage\_coordinates} presents the camera-stage coordinate frame interrelationship. The assumptions A1 and A2 of the above theorem model the design constraints for the relationship. The proof process of Theorem~\ref{THM:rel_camera_stage} is based on the properties of vectors and matrices alongside some real arithmetic reasoning.
Next, to verify the image-camera coordinate frame interrelationship, we first formalize the display resolution matrix in HOL Light as:

\vspace*{0.1cm}

\begin{mdframed}
\begin{flushleft}
\begin{definition}
\label{DEF:disp_res_mat}
\emph{Display Resolution Matrix} \\{    
\vspace*{0.1cm}
\textup{\texttt{$\vdash$ $\forall$ fx fy.\ dis\_resol\_matrix fx fy =
$\begin{bmatrix}\texttt{fx} & \hspace*{0.2cm} \texttt{0} \\ \texttt{0} & \hspace*{0.2cm} \texttt{fy} \end{bmatrix}$
}}}
\end{definition}
\end{flushleft}
\end{mdframed}

\vspace*{0.2cm}

Now, we verify the relationship between image and camera coordinate frames as:


\begin{mdframed}
\begin{flushleft}
\begin{theorem}
\label{THM:rel_image_camera}
\emph{Relationship Between Image and Camera Coordinate Frames} \\{
\textup{\texttt{$\vdash$ $\forall$ xc yc u v t fx fy.\   \\ 
$\mathtt{}$  \hspace*{0.0cm} \textbf{[A1]:}\ 0 < fx $\wedge$ \textbf{[A2]:}\ 0 < fy  \\  \vspace*{0.1cm}
$\mathtt{}$  \hspace*{0.6cm} $\Rightarrow$  \Bigg(relat\_image\_camera\_coordinates xc yc u v t fx fy $\Leftrightarrow$ $\begin{bmatrix}\texttt{u(t)} \\ \texttt{v(t)}\end{bmatrix}$ = $\begin{bmatrix}\texttt{fx $\ast$ xc(t)} \\ \texttt{fy $\ast$ yc(t)}\end{bmatrix}$ \Bigg)
}}}
\end{theorem}
\end{flushleft}
\end{mdframed}

\vspace*{0.2cm}

\noindent where the function \texttt{relat\_image\_camera\_coordinates} models the image-camera coordinate frame interrelationship. The assumptions A1 and A2 of the above theorem present the design constraints for the relationship.
Next, we formalize the transformation matrix between image and stage coordinate frames, which is used in the verification of their interrelationship:


\begin{mdframed}
\begin{flushleft}
\begin{definition}
\label{DEF:trans_mat}
\emph{Transformation Matrix} \\{
\vspace*{0.1cm}
\textup{\texttt{$\vdash$ $\forall$ fx fy alpha.\ transfor\_matrix fx fy alpha = $\begin{bmatrix}\texttt{fx $\ast$ cos (alpha)} & \hspace*{0.4cm} \texttt{fx $\ast$ sin (alpha)} \\ \texttt{-fy $\ast$ sin (alpha)} & \hspace*{0.4cm} \texttt{fy $\ast$ cos (alpha)} \end{bmatrix}$
}}}
\end{definition}
\end{flushleft}
\end{mdframed}

\vspace*{0.2cm}

Now, we verify an important relationship between the image and stage coordinate frames as:


\begin{mdframed}
\begin{flushleft}
\begin{theorem}
\label{THM:rel_image_stage}
\emph{Relationship Between Image and Stage Coordinate Frames} \\{
\textup{\texttt{$\vdash$ $\forall$ x y u v t fx fy dx dy alpha xc yc.\  \\
$\mathtt{}$ \hspace*{0.1cm} \textbf{[A1]:}\ 0 < dx $\wedge$ \textbf{[A2]:}\ 0 < dy $\wedge$ \textbf{[A3]:}\  0 < fx $\wedge$ \textbf{[A4]:}\ 0 < fy $\wedge$ \\
$\mathtt{}$ \hspace*{0.1cm}   \textbf{[A5]:}\ two\_dim\_coordinates u v t =  \\
$\mathtt{}$ \hspace*{2.5cm}   dis\_resol\_matrix fx fy $\ast\ast$ two\_dim\_coordinates xc yc t $\wedge$  \\
$\mathtt{}$ \hspace*{0.1cm} \textbf{[A6]:}\ two\_dim\_coordinates xc yc t =  \\
$\mathtt{}$ \hspace*{2.5cm} rotat\_matrix alpha $\ast\ast$ two\_dim\_coordinates x y t + displace\_vector dx dy  \\
\vspace*{0.15cm}
$\mathtt{}$ \hspace*{0.5cm} $\Rightarrow$   two\_dim\_coordinates u v t =    \\ \vspace*{0.2cm}
$\mathtt{}$ \hspace*{2.5cm} transfor\_matrix fx fy alpha $\ast\ast$ two\_dim\_coordinates x y t + $\begin{bmatrix}\texttt{fx $\ast$ dx}  \\ \texttt{fy $\ast$ dy}  \end{bmatrix}$
}}}
\end{theorem}
\end{flushleft}
\end{mdframed}

\noindent where $\mathtt{\ast\ast}$ models the matrix-vector multiplication operator in HOL Light. The assumptions A1-A4 provide the design constraints for the image-stage coordinate interrelationship. The assumption A5 provides the image-camera coordinate interrelationship. The assumption A6 presents the camera-stage coordinate interrelationship. The proof process of Theorem~\ref{THM:rel_image_stage} is mainly based on Theorems~\ref{THM:rel_camera_stage} and~\ref{THM:rel_image_camera} alongside some arithmetic reasoning on the vectors and matrices. The verification of these interrelationships ensures the correct orientation of the various important components of a robotic cell injection system, i.e., camera, working plate, microscope and the injection manipulator.

Next, we model and verify the dynamics of the robotic cell injection system. The dynamics of the $2$-DOF motion stage is based on Lagrange's equation~\cite{thorby2008structural} and mathematically represented as:


\begin{equation}\label{EQ:dyn_2_dof_stage_motion}
\begin{split}
\begin{bmatrix}m_x + m_y + m_p & \hspace*{0.3cm} 0  \\ 0 & \hspace*{0.3cm} m_y + m_p  \end{bmatrix} \begin{bmatrix} \dfrac{d^2x}{dt} \vspace*{0.3cm} \\ \dfrac{d^2y}{dt}  \end{bmatrix} + \begin{bmatrix}1 & \hspace*{0.2cm} 0  \\ 0 & \hspace*{0.2cm} 1  \end{bmatrix}  \begin{bmatrix} \dfrac{dx}{dt} \vspace*{0.3cm} \\ \dfrac{dy}{dt}  \end{bmatrix} =  \begin{bmatrix} \tau_x \vspace*{0.1cm}  \\ \tau_y  \end{bmatrix} - \begin{bmatrix} {fex}^d \vspace*{0.1cm} \\ {fey}^d  \end{bmatrix}
\end{split}
\end{equation}

\noindent where $m_x$, $m_y$ and $m_p$ represent the masses of the $xy$ positioning tables and working plate, respectively. Similarly, ${fex}^d$ and ${fey}^d$ are the $x$ and $y$ components of the desired force applied to the actuators during the robotic cell injection process, respectively. Similarly, $\tau_x$ and $\tau_y$ are the $x$ and $y$ components of the input torque of the driving motor applied during the cell injection process, respectively. We formally model Equation~(\ref{EQ:dyn_2_dof_stage_motion}) in HOL Light as:


\begin{mdframed}
\begin{flushleft}
\begin{definition}
\label{DEF:dyn_2_dof}
\emph{Dynamics of the $2$-DOF Motion Stage} \\{   
\vspace*{0.1cm}
\textup{\texttt{$\vdash$ $\forall$ mx my mp x y t taux tauy fexd feyd. \\
\hspace*{0.3cm} dynam\_2\_dof\_motion\_stage mx my mp x y t taux tauy fexd feyd $\Leftrightarrow$ \\
\hspace*{1.5cm} mass\_matrix mx my mp $\ast\ast$ sec\_order\_deriv\_stage\_coordinates x y t +  \\
\hspace*{2.3cm}  posit\_table\_matrix $\ast\ast$ fir\_order\_deriv\_stage\_coordinates x y t =    \\
\hspace*{4.8cm}  torq\_vector taux tauy - desir\_force\_vector fexd feyd
}}}
\end{definition}
\end{flushleft}
\end{mdframed}

\vspace*{0.2cm}

\noindent where \texttt{mass\_matrix} is the matrix containing the respective masses, i.e., $m_x$, $m_y$ and $m_p$, and \texttt{posit\_table\_matrix} is a diagonal matrix. Similarly, \texttt{desir\_force\_vector}  and \texttt{torq\_vector} are the desired force and the applied torque vectors, respectively, i.e., the elements of these vectors represent the components of the desired force and applied torque. The functions \texttt{fir\_order\_deriv\_stage\_coordinates} and \texttt{sec\_order\_deriv\_stage\_coordinates} represent the vectors containing first-order and second-order derivatives of the stage coordinates, respectively:


\begin{mdframed}
\begin{flushleft}
\begin{definition}
\label{DEF:fst_n_scd_ord_der_vec}
\emph{Vectors Containing First and Second-order Derivatives of the Stage Coordinates} \\{   
\vspace*{0.1cm}
\textup{\texttt{$\vdash$ $\forall$ x y t.\ fir\_order\_deriv\_stage\_coordinates x y t = deriv\_vector\_first [x; y] t \\
$\vdash$ $\forall$ x y t.\ sec\_order\_deriv\_stage\_coordinates x y t = deriv\_vector\_second [x; y] t
}}}
\end{definition}
\end{flushleft}
\end{mdframed}

\vspace*{0.2cm}

\noindent where \texttt{deriv\_vector\_first} and \texttt{deriv\_vector\_second} take a list containing the functions of data-type $\mathds{R} \rightarrow \mathds{R}$ and output the vectors containing the corresponding first and second-order derivatives of the functions.~\cite{adnan17robcellinjsystp}.

If the desired force and the applied torque vectors are zero, then the injection pipette does not touch the cells. Thus, for this particular case, Equation~(\ref{EQ:dyn_2_dof_stage_motion}) can be expressed as:


\begin{equation}\label{EQ:dyn_2_dof_stage_motion_homog}
\begin{split}
\begin{bmatrix}m_x + m_y + m_p &  \hspace*{0.3cm} 0  \\ 0 & \hspace*{0.3cm} m_y + m_p  \end{bmatrix} \begin{bmatrix} \dfrac{d^2x}{dt} \vspace*{0.3cm} \\ \dfrac{d^2y}{dt}  \end{bmatrix} + \begin{bmatrix}1 & \hspace*{0.2cm} 0  \\ 0 & \hspace*{0.2cm} 1  \end{bmatrix}
\begin{bmatrix} \dfrac{dx}{dt} \vspace*{0.3cm} \\ \dfrac{dy}{dt}  \end{bmatrix} = \begin{bmatrix} 0  \\ 0  \end{bmatrix} \end{split}
\end{equation}

We verify the solution of the above equation in HOL Light as the following theorem:


\begin{mdframed}
\begin{flushleft}
\begin{theorem}
\label{THM:soln_ver_dyn_stage_motion_homog}
\emph{Verification of Solution of Dynamical Behaviour of Motion Stage} \\{
\textup{\texttt{$\vdash$ $\forall$ x y mx my mp taux tauy fexd feyd alpha x0 y0 xd0 yd0.\ \\
$\mathtt{}$ \hspace*{0.1cm} \textbf{[A1]:}\ 0 < mx  $\wedge$ \textbf{[A2]:}\ 0 < my  $\wedge$ \textbf{[A3]:}\ 0 < mp $\wedge$  \\
\vspace*{0.1cm}
$\mathtt{}$ \hspace*{0.1cm} \textbf{[A4]:}\ x(0) = x0 $\wedge$ \textbf{[A5]:}\ y(0) = y0 $\wedge$ \textbf{[A6]:}\ $\mathtt{\dfrac{dx}{dt}}$(0)= xd0 $\wedge$ \textbf{[A7]:}\ $\mathtt{\dfrac{dy}{dt}}$(0)= yd0 $\wedge$  \\ \vspace*{0.2cm}
$\mathtt{}$ \hspace*{0.1cm}  \textbf{[A8]:}\ $\mathtt{}$   $\begin{bmatrix}\texttt{fexd}  \\ \texttt{feyd}  \end{bmatrix}$ = $\begin{bmatrix}\texttt{0}  \\ \texttt{0}  \end{bmatrix}$ $\wedge$ \textbf{[A9]:}\ $\mathtt{}$   $\begin{bmatrix}\texttt{taux}  \\ \texttt{tauy}  \end{bmatrix}$ = $\begin{bmatrix}\texttt{0}  \\ \texttt{0}  \end{bmatrix}$ $\wedge$  \\ \vspace*{0.1cm}
$\mathtt{}$ \hspace*{0.1cm}  \textbf{[A10]:}\ ($\forall$ t. x(t) = (x0 + xd0 $\ast$ (mx + my + mp))  \\  \vspace*{0.1cm}
$\mathtt{}$  \hspace*{4.0cm}  - xd0 $\ast$ (mx + my + mp) $\ast$  \large$\mathtt{e^{{\frac{-1}{mx + my + mp}} t}}$ $\wedge$  \\  \vspace*{0.1cm}
$\mathtt{}$ \hspace*{0.05cm}  \textbf{[A11]:}\ ($\forall$ t. y(t) = (y0 + yd0 $\ast$ (my + mp))  \\ \vspace*{0.2cm}
$\mathtt{}$  \hspace*{4.0cm}  - yd0 $\ast$ (my + mp) $\ast$  \large{$\mathtt{e^{{\frac{-1}{my + mp}} t}}$}   \\ \vspace*{0.1cm}
$\mathtt{}$  \hspace*{1.0cm} $\Rightarrow$ dynam\_2\_dof\_motion\_stage mx my mp x y t taux tauy fexd feyd
}}}
\end{theorem}
\end{flushleft}
\end{mdframed}

\noindent  The assumptions A1-A3 provide the conditions on the masses \texttt{mx}, \texttt{my} and \texttt{mp}, i.e., all masses are positive. The assumptions A4-A7 present the values of coordinates \texttt{x} and \texttt{y} and their first-order derivatives $\mathtt{\frac{dx}{dt}}$ and $\mathtt{\frac{dy}{dt}}$ at $t = 0$. The assumptions A8-A9 model the constraints on the components of the desired force and the torque, respectively, i.e., the desired force and torque vectors are zero. The assumptions A10-A11 present the values of $xy$ coordinates at any time $t$. Finally, the conclusion provides the dynamical behaviour of the $2$-DOF motion stage. The verification of Theorem~\ref{THM:soln_ver_dyn_stage_motion_homog} is mainly based on the properties of real derivatives, transcendental functions, vectors and matrices. Next, we verify an alternate representation of the image-stage coordinate interrelationship, which depends on the dynamical behaviour of the motion stage (Definition~\ref{DEF:dyn_2_dof}) and is an important property for analyzing cell injection systems. For this purpose, we first formalize the inertia and positioning table matrices:


\begin{mdframed}
\begin{flushleft}
\begin{definition}
\label{DEF:pos_tab_n_iner_mat}
\emph{Inertia and Positioning Table Matrices} \\{
\textup{\texttt{$\vdash$ $\forall$ mx my mp fx fy alpha.\ inertia\_matrix mx my mp fx fy alpha = \\
$\mathtt{}$ \hspace*{3.0cm} mass\_matrix mx my mp $\ast\ast$ matrix\_inv (transfor\_matrix fx fy alpha) \\
$\mathtt{}$$\vdash$ $\forall$ fx fy alpha.\ posit\_table\_matrix\_fin fx fy alpha =    \\
$\mathtt{}$ \hspace*{3.0cm} posit\_table\_matrix $\ast\ast$ matrix\_inv (transfor\_matrix fx fy alpha)
}}}
\end{definition}
\end{flushleft}
\end{mdframed}


\noindent where the HOL Light function \texttt{matrix\_inv} takes a matrix \texttt{A:}${\mathds{R}^M}^N$ and returns its inverse ($A^{-1}$). Now, we verify the alternate form of the relationship between image and stage coordinates in HOL Light:


\begin{mdframed}
\begin{flushleft}
\begin{theorem}
\label{THM:rel_image_stage_alt}
\emph{Alternate Representation of the Image-Stage Coordinate Interrelationship} \\{
\textup{\texttt{$\vdash$ $\forall$ xc yc u v x y fx fy dx dy mx my mp taux tauy fexd feyd alpha.\   \\ \vspace*{0.05cm}
$\mathtt{}$ \hspace*{-0.2cm} \textbf{[A1]:}\ 0 < dx  $\wedge$ \textbf{[A2]:}\ 0 < dy  $\wedge$  \textbf{[A3]:}\ 0 < fx  $\wedge$ \textbf{[A4]:}\ 0 < fy  $\wedge$  \\
$\mathtt{}$ \hspace*{-0.2cm} \textbf{[A5]:}\ invertible (transfor\_matrix fx fy alpha) $\wedge$  \\
$\mathtt{}$ \hspace*{0.0cm}\textbf{[A6]:}\ ($\forall$ t.\ u real\_differentiable atreal t) $\wedge$   \\ 
$\mathtt{}$ \hspace*{0.0cm}\textbf{[A7]:}\ ($\forall$ t.\ v real\_differentiable atreal t) $\wedge$   \\ \vspace*{0.1cm}
$\mathtt{}$ \hspace*{0.0cm}\textbf{[A8]:}\ ($\forall$ t.\ $\mathtt{\dfrac{du}{dt}}$ real\_differentiable atreal t) $\wedge$   \\ \vspace*{0.1cm}
$\mathtt{}$ \hspace*{0.0cm}\textbf{[A9]:}\ ($\forall$ t.\ $\mathtt{\dfrac{dv}{dt}}$ real\_differentiable atreal t) $\wedge$   \\ \vspace*{0.1cm}
$\mathtt{}$ \hspace*{0.0cm}\textbf{[A10]:}\ ($\forall$ t.\ relat\_image\_camera\_coordinates xc yc u v t fx fy) $\wedge$  \\
$\mathtt{}$ \hspace*{0.0cm}\textbf{[A11]:}\ ($\forall$ t.\ relat\_camera\_stage\_coordinates xc yc x y alpha dx dy t) $\wedge$  \\
$\mathtt{}$ \hspace*{0.0cm}\textbf{[A12]:}\ dynam\_2\_dof\_motion\_stage mx my mp x y t taux tauy fexd feyd   \\
$\mathtt{}$   \hspace*{0.2cm}  $\Rightarrow$ inertia\_matrix mx my mp fx fy alpha $\ast\ast$ sec\_order\_deriv\_image\_coordinates u v t +  \\
$\mathtt{}$   \hspace*{1.5cm}   posit\_table\_matrix\_fin fx fy alpha $\ast\ast$ fir\_order\_deriv\_image\_coordinates u v t =  \\
$\mathtt{}$   \hspace*{3.0cm}   torq\_vector taux tauy - desir\_force\_vector fexd feyd
}}}
\end{theorem}
\end{flushleft}
\end{mdframed}


The assumptions A1-A4 present the design constraints for the relationship between the image and stage coordinates. The assumption A5 describes the invertibility of the transformation matrix \texttt{transf\_mat}, i.e., existence of its inverse. The assumptions A6-A9 provide the differentiability conditions for the image coordinate and their first-order derivatives.
The assumptions A10-A11 represent the image-camera and camera-stage coordinates interrelationships. The assumption A12 provides the dynamical behavior of the $2$-DOF motion stage. Finally, the conclusion of Theorem~\ref{THM:rel_image_stage_alt} presents the alternate form of the relationship between the image and stage coordinate frames. The proof process of the theorem is based on the properties of the real derivative, vectors and matrices alongwith some real arithmetic reasoning.

Due to the undecidable nature of the higher-order logic, the formalization presented in Section~\ref{SEC:form_analy_cell_inj_sys}, involved manual interventions and human guidance.
The proof effort involved $520$ lines-of-code and $12$ man-hours. The details about
the reported formalization can be found in our proof script~\cite{adnan17robcellinjsystp}. The distinguishing feature of our formalization is that all the verified theorems are universally quantified and can thus be specialized to the required values based on the requirement of the analysis of the cell injection systems. Moreover, our higher-order logic based approach allows us to model the dynamical behaviour of the robotic cell injection systems involving derivatives (Equation~(\ref{EQ:dyn_2_dof_stage_motion})) in their true form, whereas, they are discretized and modeled using a state-transition system in their model checking based analysis~\cite{sardar2017towards}, which may compromise the correctness and completeness of the corresponding analysis.

\section{Conclusion}\label{SEC:conclusion}

In this paper, we proposed formal modeling of robotic cell injection systems in higher-order logic. We first formalized the camera, image and stage coordinate frames, which are the vital components of a robotic cell injection system, and formally verified their interrelationship using the HOL Light proof assistant. We also formalized the dynamical behaviour of the $2$-DOF motion stage based on differential equations and verified their solutions in HOL Light.

We have already extended the reported formalization by formalizing the impedance force control and image-based torque controller, which are mainly responsible for the process of cell injection~\cite{rashid2018formalrobotics}. In future, we plan to verify the relationship between both of these controllers. Another future direction is to formalize the $3$-DOF motion stage and formally analyze the dynamics of the corresponding robotic cell injection system.

%

\bibliographystyle{alpha}
\bibliography{biblio}

%
%
%

\end{document}